# SIRT6 knockout cells resist apoptosis initiation but not progression:

# A computational method to evaluate the progression of apoptosis

Sergii Domanskyi,[1] Justin W. Nicholatos,[2] Joshua E. Schilling,[1]
Vladimir Privman,[1] Sergiy Libert[2]*

[1]Department of Physics, Clarkson University, Potsdam, NY 13676

[2]Department of Biomedical Sciences, Cornell University, Ithaca, NY 14853

*libert@cornell.edu

## Abstract

Apoptosis is essential for numerous processes, such as development, resistance to infections, and suppression of tumorigenesis. Here, we investigate the influence of the nutrient sensing and longevity-assuring enzyme SIRT6 on the dynamics of apoptosis triggered by serum starvation. Specifically, we characterize the progression of apoptosis in wild type and SIRT6 deficient mouse embryonic fibroblasts using time-lapse flow cytometry and computational modelling based on rate-equations and cell distribution analysis.

We find that SIRT6 deficient cells resist apoptosis by delaying its initiation. Interestingly, once apoptosis is initiated, the rate of its progression is higher in SIRT6 null cells compared to identically cultured wild type cells. However, SIRT6 null cells succumb to apoptosis more slowly, not only in response to nutrient deprivation but also in response to other stresses. Our data suggest that SIRT6 plays a role in several distinct steps of apoptosis. Overall, we demonstrate the utility of our computational model to describe stages of apoptosis progression and the integrity of the cellular membrane. Such measurements will be useful in a broad range of biological applications.

## Significance Statement

We describe a computational method to evaluate the progression of apoptosis through different stages. Using this method, we describe how cells devoid of SIRT6 longevity gene respond to apoptosis stimuli, specifically, how they respond to starvation. We find that SIRT6 cells resist apoptosis initiation; however, once initiated, they progress through the apoptosis at a faster rate. These data are first of the kind and suggest that SIRT6 activities might play different roles at different stages of apoptosis. The model that we propose can be used to quantitatively evaluate progression of apoptosis induced by numerous treatments and genetic manipulations and will be useful in studies of cancer treatments, development, degenerative disorders, and other areas, where apoptosis is involved.

## Introduction

The capacity of cells to undergo programmed cell death or apoptosis is critical for numerous organismal functions, which range from development, to immunocompetence, to cancer resistance [1]. Dysregulation of apoptosis often leads to devastating diseases. For example, overactive apoptosis in neurons is the cause of degenerative diseases, such as Alzheimer's [2] and Parkinson's [3]; insufficient apoptosis is the cause for the production of self-reactive immune cells, which in turn can cause numerous autoimmune disorders [4], such as type I diabetes. Likewise, suppression of apoptosis is a key step in cancer initiation and progression, and p53—a master regulator of apoptosis—is the most commonly mutated gene in all human cancers [5]. Interestingly, it has been reported that the metabolic enzyme SIRT6 is also deleted in over 35% of cancers [6], suggesting the role of this gene in controlling programmed cell death.

Sirtuins are a family of NAD-dependent enzymes which sense nutrient availability and adjust metabolic pathways to deal with possible changes [7]. This property of sirtuin enzymes had been proposed to mediate the ability of calorie restriction to extend longevity in numerous organisms, including single cell yeast [8], fruit flies [9], mice, and monkeys [10]. Interestingly, one member of this family, SIRT6 is shown to extend longevity of mice when overexpressed [11], and genetic variants of this gene have been associated with the lifespan in humans [12]. SIRT6 possesses at least three distinct enzymatic activities, and one of its suggested functions is regulation of apoptosis [13,14], details of which we explore here.

The pathways that initiate and ensure progression of apoptosis are reasonably well documented [15]. However, the degree of influence of different factors that contribute to apoptosis is not known, and often described only qualitatively. To understand the dynamics of apoptosis, and more importantly the influence of nutrients and SIRT6 on such dynamics, we developed a computational model, which when paired with time-course experiments, allows us to extract numerical parameters associated with cell death. Specifically, we can evaluate the rates of apoptosis phases; these include sensitization to, initiation and progression of, as well as certain aspects of heterogeneity of cells at different stages of apoptosis. Information on how specific genes and/or therapeutics influence these parameters will aid our understanding of numerous disorders, for which dysregulation of apoptosis is a culprit. In this manuscript, we utilize our numerical model to quantitatively describe the impact of SIRT6 on nutrient-stress-induced apoptosis.

## Results

To test the impact of SIRT6 on apoptosis *in vitro*, we used immortalized mouse embryonic fibroblasts (MEFs) derived either from SIRT6 knockout (KO) mice (Fig. 1A) or from their wild type littermates. We stressed these cells with a variety of insults, after which cells were stained with propidium iodide (PI) and Annexin V, and analyzed using flow cytometry to detect and quantify necrosis and apoptosis. We found that cells lacking SIRT6 resist apoptosis following a diverse array of insults (Fig. 1C-D). When we challenged cells with rotenone, which inhibits complex I of the electron transport chain of mitochondria, 19% of wild type cells underwent apoptosis, while only 9% of SIRT6 KO cells did (Fig. 1C, 53% reduction of apoptosis). When we challenged cells with the proteasome inhibitor MG-132, 41% of wild type cells underwent apoptosis, compared to 12% of SIRT6 KO (Fig. 1D, 71% reduction of apoptosis). When we challenged cells with DNA damaging agent—etoposide, 41% of wild type cells underwent apoptosis, compared to 11% of SIRT6 KO (Fig. 1F, 72% reduction of apoptosis). By far the largest difference in survival between the SIRT6 KO and WT cells was observed with serum starvation stress. WT MEFs displayed 83% greater cell death than their KO counterparts (Fig. 1E). Similar results were obtained among three independent batches of WT and SIRT6 KO primary cells collected from different mice. These results reinforce the central role of SIRT6 in nutrient sensing and apoptosis promotion, which warrants more detailed investigation.

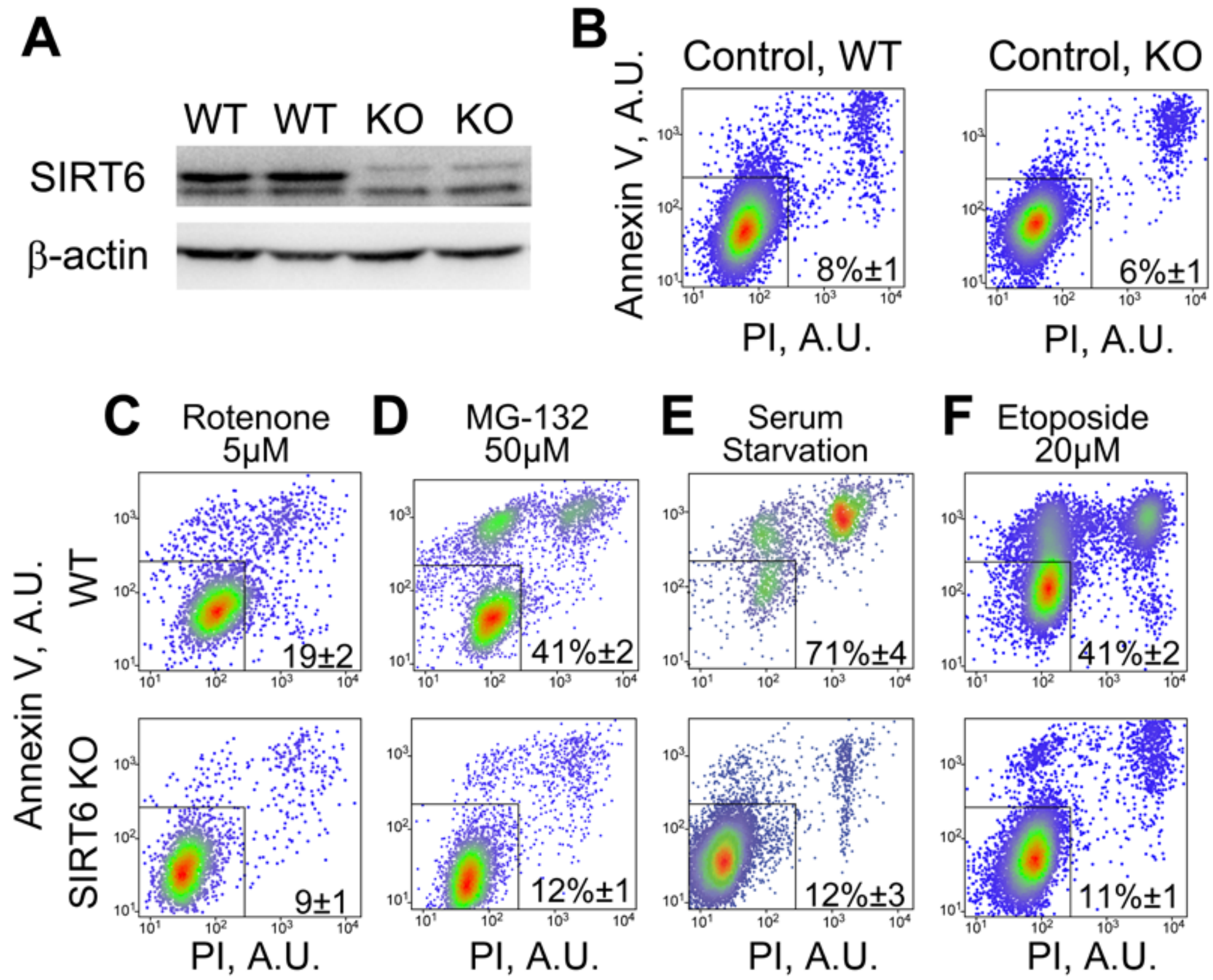

**Figure 1.** SIRT6 knockout cells are resistant to apoptosis. **(A)** SDS-PAGE analysis of SIRT6 abundance in WT and SIRT6 KO cells. **(B-F)** Representative flow cytometry analysis profiles are shown for WT and SIRT6 knockout MEFs. Annexin V-FITC conjugated antibodies and propidium iodide staining were used to identify apoptotic and necrotic cells. Fluorescence intensities are plotted on a log scale. Each dot represents a single cell. Dot coloring reflects local cell density in the given area of the graph. A.U. – arbitrary units of fluorescence intensity. Gating and bootstrap analysis were used to compute the mean fraction of dead or dying cells and the confidence interval. Prior to analysis, cells were stressed with **(C)** rotenone, **(D)** MG-132 proteasome inhibitor, **(E)** serum starvation, **(F)** etoposide, or **(B)** treated with vehicle as a control. In all the cases, WT cells were more susceptible to apoptosis then SIRT6 KO cells.

To study the impact of serum withdrawal and SIRT6 on the initiation and progression of apoptosis, we performed time-lapse experiments. SIRT6 KO or WT cells were subjected to one of the two treatments: normal tissue culture conditions or culture without serum for 72 hours. Every 12 hours, a subset of cells was collected, stained with PI and Annexin V, and analyzed using flow cytometry. At least 2000 cells were analyzed for every measurement time for each treatment (Fig. 2). Notice an appearance of the pre-apoptotic and apoptotic populations of cells (cell populations positive for PI and Annexin V) as time progresses (Fig. 2). The difference between SIRT6 knockout cells and WT can be clearly seen at 72 hours, when almost 60% of WT cells are undergoing apoptosis versus only 12% of knockout cells.

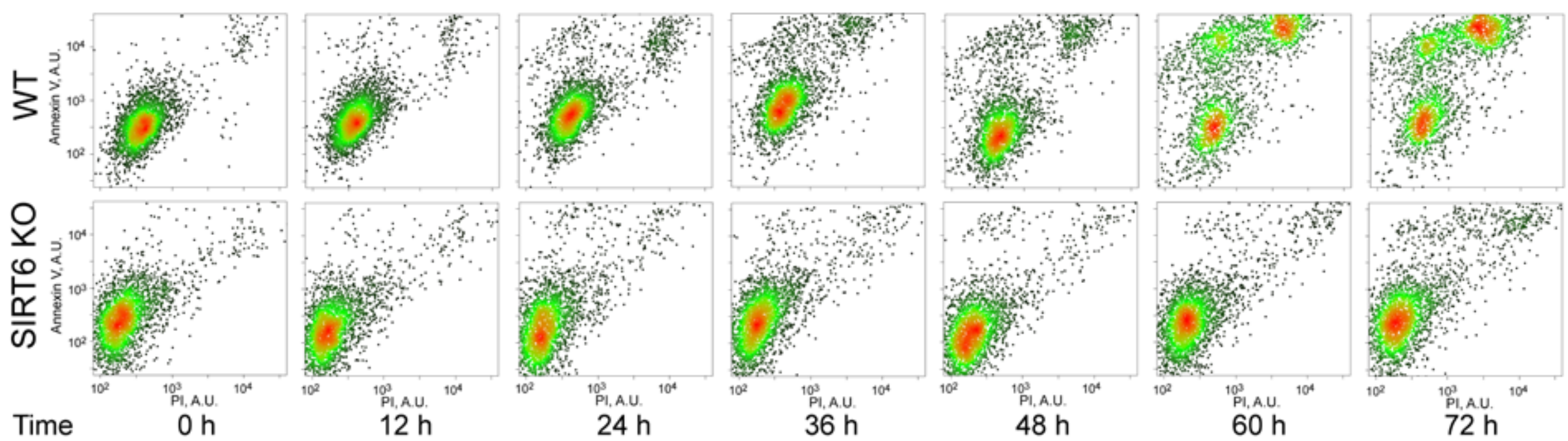

**Figure 2.** Dynamic progression of serum starvation induced apoptosis. Flow cytometry analysis profiles are shown for WT and SIRT6 knockout MEFs at 12-hour intervals. Annexin V-FITC conjugated antibodies and propidium iodide staining were used to identify apoptotic and necrotic cells. Fluorescence intensity is plotted on a log scale. Each dot represents a single cell. Dot coloring reflects local cell density in the given area of the graph. A.U. – arbitrary units of fluorescence intensity.

Staining with PI and Annexin V is the most common method used to identify apoptotic cells [16]. Cells with negative stain for both dyes are considered normal (healthy). As apoptosis initiates, mitochondria release cytochrome C, which activates signaling cascades that mobilize proteases and nucleases, ultimately resulting in cells losing energy and membrane rigidity. Consequent inversions of the plasma membrane can be detected by Annexin V stain. Cells positive for Annexin V but negative for PI are at the early stage of apoptosis and we identify such cells as “pre-apoptotic”. As apoptosis progresses, loss of the cellular membrane and the nuclear envelope integrity allows influx of PI, which fluoresces when bound to DNA. Cells positive for both Annexin V and PI are identified as “apoptotic”. Rapid PI influx can also be caused by necrosis, therefore cells positive for PI, but negative for Annexin V are identified as “necrotic”. As can be seen on flow cytometry plots (Figs. 1, 2), cells undergoing apoptosis have a distribution of staining intensities for both dyes. More importantly, there is a correlation between intensities (distributions tend to be diagonally elongated with respect to the X-, Y-axis), which is a phenomena we discuss later in the context of our analysis of staining distributions.

To model the initiation and progression of apoptosis in MEFs, we consider the known molecular pathways of apoptosis progression (Fig. 3A). In normal conditions, cells receive pro-survival signals through various growth factor receptors, such as insulin receptor or growth hormone receptor. We designate fractions of such cells as $C_1$. As growth factor signaling decreases due to serum withdrawal, pro-survival signaling, such as classical growth kinases: AKT, PI3K, and ERK would decrease their activity, which puts cells into still normal, but apoptosis “sensitive” state, which we designate as $C_2$. The strength of growth signaling might vary from cell to cell, and in cases where signaling is strong enough, cells can divide, to produce two daughter cells. However, if signaling is sufficiently low, it will allow accumulation of early pro-apoptotic factors, such as Bad, BAX, and tBID, which will cause mitochondrial release of Cytochrome C and AIF (apoptosis induction factor). This cellular state, we designate as pre-apoptotic or $A$. We expect that these cells begin accumulating Annexin V staining, as activation of caspases causes destabilization of the cytoskeleton and thus the cell membrane. Finally, when caspases, such as caspase-3 and caspase-9 are fully activated, DNA fragmentation and cytoskeleton destruction will ultimately kill the cell. We expect these cells to be positively stained by both Annexin V and PI, and we designate the fraction of such cells as $D$. Additionally, cells can undergo necrosis, which we designate as $N$. Finally, the transition of cells from one state to another happens at certain rates. For example, we designate the rate of transition from $C_1$ to $C_2$ as $S$; and from $A$ to $D$ as $G$, etc. Using experimental data for the dynamics of apoptosis (Fig. 2), we determined the fractions of normal: $C$ (equal $C_1 + C_2$), pre-apoptotic: $A$, apoptotic (dead by apoptosis): $D$, and necrotic: $N$ cells as the functions of time, $t$. For this, each fixed-time snapshot dataset was divided into four quadrants at PI 2000 A.U. and Annexin 4000 A.U. to yield the mentioned fractions of cells. These fractions were further adjusted by subtracting the background, selected to yield 100% normal cells at time zero. The quadrant cutoffs were selected for clear differentiation of the cell types for serum-starved WT cells at times past 60 hours, because for these cells at later times the differences between the various cell-type distributions were defined the best.

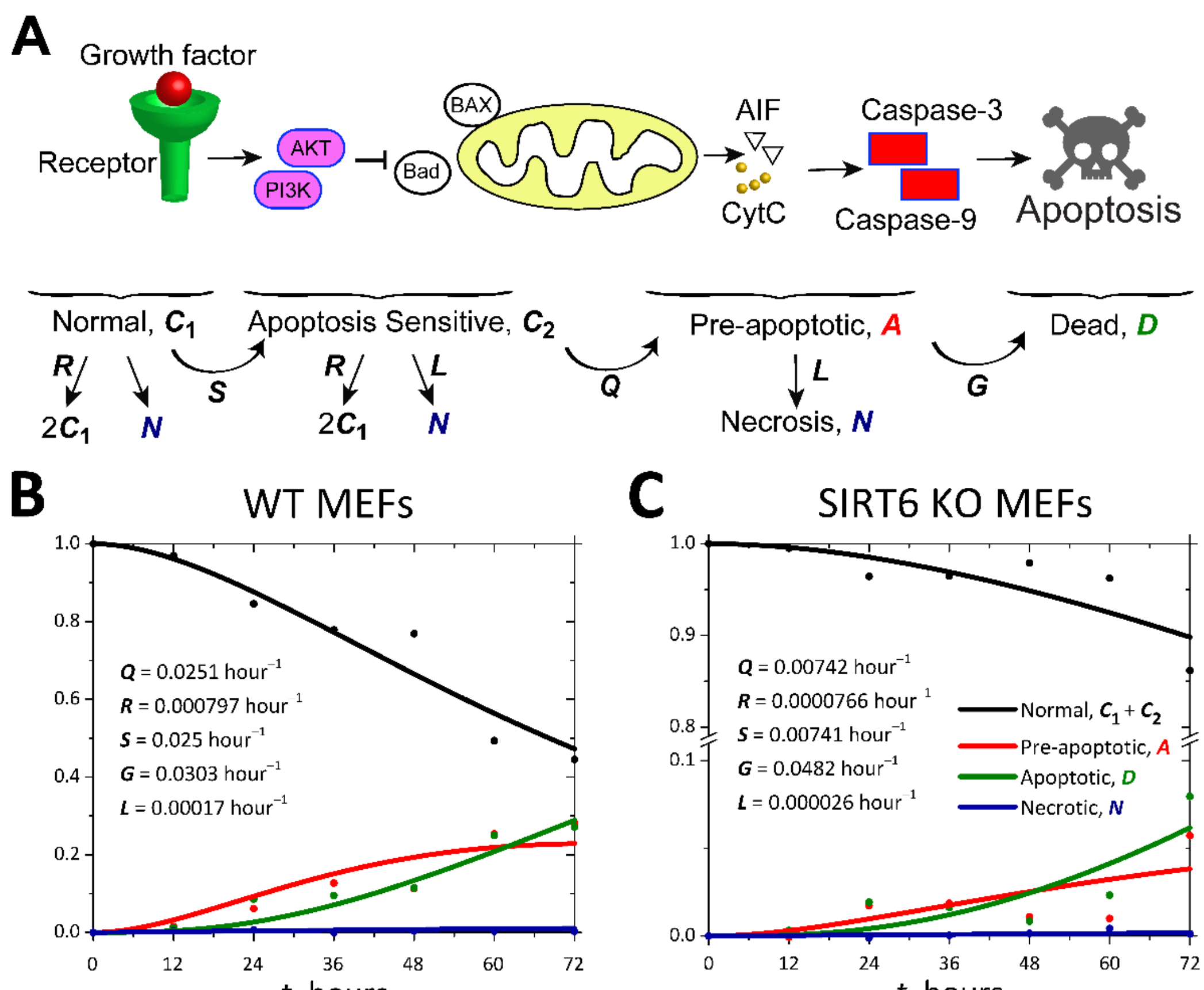

**Figure 3. (A)** Simplified schematics for the apoptosis initiation and progression. To model the progression of apoptosis, we defined cellular states ($C_1$, $C_2$, $A$, $D$, $N$), which correspond to different stages of apoptosis (described in the text). The cell fractions of the types $A$, $C$, $D$, $N$ at different time points were extracted from experimental data (Fig. 2). Rate-equation modeling allows the extraction of rates, such as the rate of apoptosis initiation or the rate of apoptosis progression, all of which define the impact of manipulation (genetic or environmental, in our case SIRT6 mutation) on apoptosis dynamics. **(B,C)** Data (circles) and numerical model fits (lines, color-coded to the data) for various types of **(B)** serum-starved WT cells; **(C)** serum-starved SIRT6 KO cells. The values of the rates that allow best data fits are given in the panels.

The cellular state transition rates define the progression of apoptosis, and serve as a quantitative measurement for the strength of apoptosis stimulus and strength of intervention (genetic or environmental) that modifies cellular apoptosis. To calculate these rates, we developed a rate-equation model that describes the included processes as functions of time (Fig. 3). The measured data provide the time dependence of the total count of normal cells, $C(t)$. However, it is known that with time cells become more susceptible to initiating apoptosis (Fig. 3A), and, interestingly, we found that setting up rate equations only for cell fractions $C$, $A$, $D$, and $N$, without accounting for this property is not sufficient to fit the experimental data. Indeed, data fitting necessitates designating the fraction of the cells that are still well resistant to becoming pre-apoptotic by $C_1$, whereas the rest of the normal cells' fraction,

$$C_2 = C - C_1 \ , \qquad (1)$$

are cells susceptible to apoptosis initiation.

The set of rate equations used to calculate cell transitions through stages of apoptosis are:

$$\frac{dC_1}{dt} = R(C_1 + 2C_2) - SC_1 - LC_1 \ , \qquad (2)$$

$$\frac{dC_2}{dt} = -RC_2 + SC_1 - QC_2 - LC_2 \ , \qquad (3)$$

$$\frac{dA}{dt} = QC_2 - GA - LA \ , \qquad (4)$$

$$\frac{dD}{dt} = GA \ , \qquad (5)$$

$$\frac{dN}{dt} = L(C_1 + C_2 + A) \ . \qquad (6)$$

Here all normal cells divide with the rate $R$, and furthermore, we assume that the produced daughter cells are both in the $C_1$ population. We also assume that $C_2$ cells (but not $C_1$ cells) initiate apoptosis at the rate $Q$; the $C_1$ cells join the $C_2$ population at the rate $S$; the rate of cellular necrosis is $L$; and the rate of apoptosis progression is $G$ (the rates are schematically depicted on Fig. 3A).

The rate-constant values for the set of equations (1-6) were obtained by least-squares fit of the data for the mentioned cell-type fractions. The results are given in Figs. 3B, 3C. The rate $G$ was estimated independently to a good precision, because its determination depends on the population of pre-apoptotic cells. The pre-apoptotic cells are produced by the succession of processes with the rates $Q$, $R$ and $S$, which makes the numerically fitted values of these three rates somewhat interdependent. Out of these, we found that the quality of the fit for the rate $S$ is the best, whereas the rate $R$ is determined the least precisely. The rate $L$ of necrosis is not determined precisely here, due to the small fraction of such cells and the noise in the data. Its value is weakly correlated with the rate $G$, as the latter determines how many live cells (normal and pre-apoptotic) remain.

It is noteworthy that the values of $Q$, $G$, $L$ are comparable to estimates obtained in our previous study [17] that used a similar numerical model to describe chicken fibroblast death due to Bursal disease virus infection. Specifically, the value for the rate of apoptosis progression, $G = 0.0303 \ \mathrm{hour}^{-1}$, fitted for the present data for WT serum-starved cells is close to that in [17], $G = 0.0231 \ \mathrm{hour}^{-1}$, suggesting some sort of "universality" of the rates of cell-death processes. The rate of apoptosis initiation (rate $Q$, Fig. 3B) is also close to that of "highly-infected" cells [17].

## Discussion

Elucidating and quantifying the influence of SIRT6 on apoptosis dynamics is an example of what can be discovered using our model. Comparison of the rates that define apoptosis progression in WT and SIRT6 KO cells allows us to define the function of SIRT6 in this process. We find that the rate with which cells become sensitive to apoptosis (*S*) and the rate of apoptosis initiation (*Q*) is ~3 times slower for the cells that lack SIRT6. However, when apoptosis is initiated, the rate of apoptosis progression (*G*) is ~60% greater for the cells that lack SIRT6. These observations are intriguing, as they suggest that SIRT6 is involved in several distinct steps of apoptosis progression.

The rate of cell division (fitted parameter *R*) is much smaller than a typical value for cells not subject to stresses [17]. This is consistent with the observation that growth signaling provided by fetal bovine serum is necessary for rapid cell division [18]. Interestingly, we find that in the state of serum withdrawal, this rate is significantly higher in WT cells, which independently confirms that part of the SIRT6 function is to alter cellular metabolism in response to changing nutrient availability.

The intensity of cell staining by PI and Annexin V varies between individual cells, and the distribution generated by this variability can be analyzed further to extract additional interesting information (Fig. 4). An example of the fitted cell distributions by their degree of staining is shown (Fig. 4A, 4B) for both the WT and SIRT6 KO cells after 72 hours of serum starvation. For wild type cells under serum starvation, normal, pre-apoptotic, and apoptotic groups of cells and their distributions are well defined. For example, the group of cells designated as necrotic (only positive for PI) constitutes only 1.1% of the total. The green dots used for a "bar-chart"-like representation of the data (Fig. 4A, 4B) bin the fraction of the cells that have the amounts of two dyes in a square bin of the size $h = \log_{10}(10^4\ \text{A.U.})/50$. Cells outside of the $(10\ \text{A.U.} - 10^5\ \text{A.U.})$ range were excluded from the analysis. The fitted peaked distributions (Fig. 4) were obtained as follows. For each of the cell groups separately, random-noise deviation from the average values, $(x_0, y_0)$, was assumed, but with possible correlation in the distributions along the logarithmic "coordinates" $x = \log_{10}(\text{PI, A.U.})$ and $y = \log_{10}(\text{Annexin V, A.U.})$. Thus, the set of values for the staining within each cell type was fitted to the probability distribution that is Gaussian in two variables, but rotated angle $\theta$ in the $(x, y)$ plane,

$$F(x,y) = \frac{1}{2\pi ab} e^{-[(x-x_0)\cos\theta+(y-y_0)\sin\theta]^2/(2a^2)-[(y-y_0)\cos\theta-(x-x_0)\sin\theta]^2/(2b^2)} . \qquad (7)$$

The peaks were normalized to the fractions of the corresponding cells.

Note that the expression describing the distributions, equation (7) depends on the fitted parameter $\theta$. This parameter in indicative of the correlation between the processes of staining by different dyes, and interestingly, this angle is different for cells at different stages of apoptosis. To illustrate this further, we plotted the cross-section of fitted distributions, see Fig. 4C. The significance of the angle of rotation, $\theta$, is that off-axes angles, e.g., $\theta \simeq 45°$, indicate a correlation in the attachment of the two dyes. The attachment of both PI and Annexin V is limited by the integrity of the cell membrane, therefore some correlation is expected. Specifically, PI fluoresces when it is bound to DNA (though its delivery is also limited by the membrane permeability), and Annexin fluoresces when attached to the inverted parts of the (damaged) membrane. Correlations, therefore, would provide information about membrane mechanics of dying cells. Such information is generally of great interest in aging and Alzheimer's disease research, since Aβ peptides are proposed to induce death by damaging membranes of neurons [19].

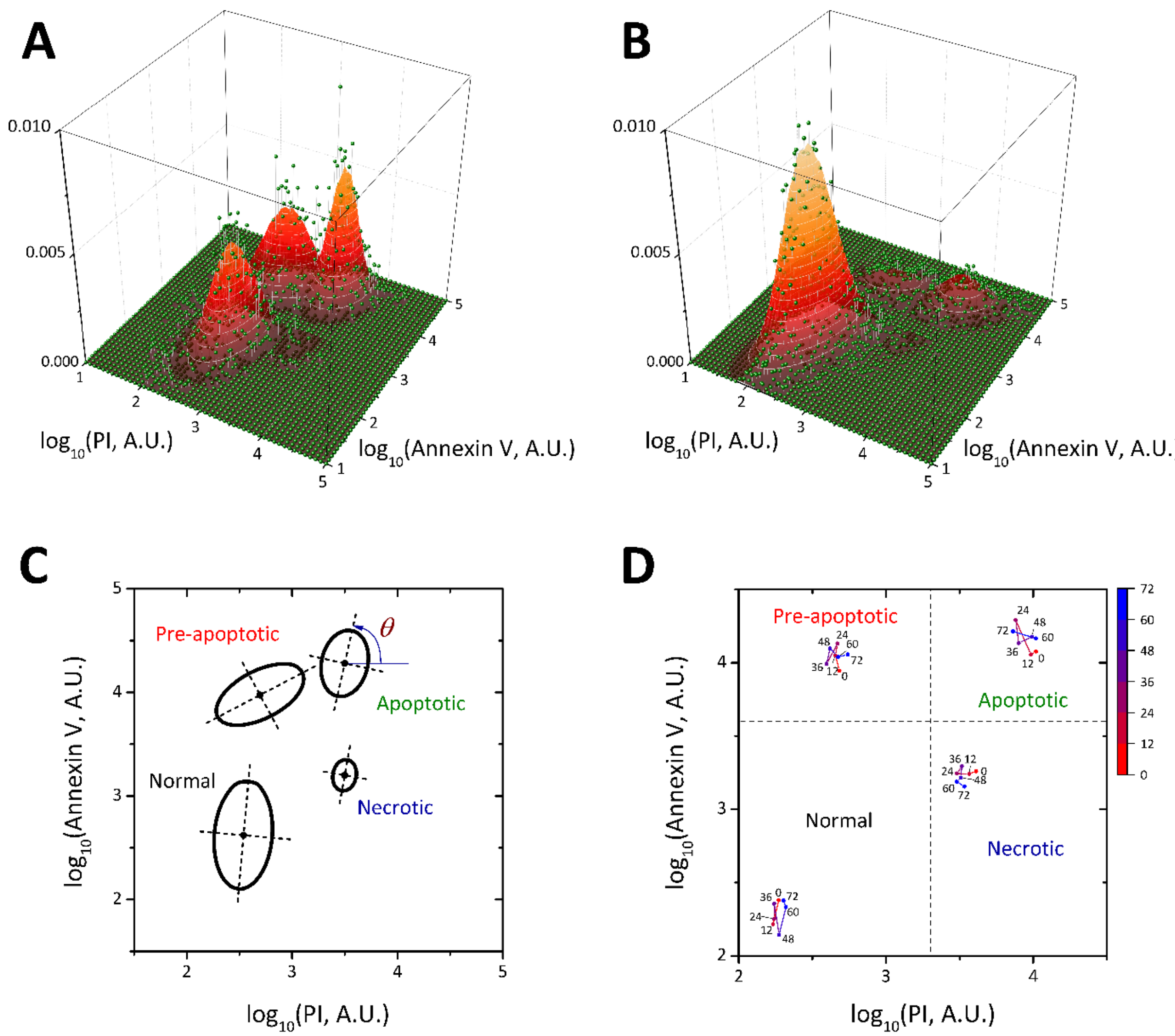

**Figure 4.** Fitted cell distributions by their degree of staining within each of the cell-type groups after 72 hours of serum starvation, for (**A**) wild type cells, and (**B**) SIRT6 knockout cells. Each distribution is of the assumed form, equation (7), and is normalized to integrate to the fraction of that type of cells. The bin designation for the "bar chart"-style dot-plot (green dots) is explained in the text. **(C)** Cross-sections of the peaks shown in **(A)** at heights that are 1/2 of the maximum values for each of the distributions for the WT cells after 72 hours of serum starvation. Notice the different angle $\theta$ for cells at different stages of apoptosis. **(D)** Positions of the centers of the peaks for each of the distributions (calculated without subtracting the background, for this panel only) for the SIRT6 KO cells, for varying times of serum starvation (marked in the panel). The progression of apoptosis is depicted by connecting the points according to the color-coding shown in the bar on the right, with the corresponding times of measurement indicated by numbers.

We further investigated the behavior of locations of the cell staining distributions during apoptosis progression. Figure 4D demonstrates "travel" of the center of the computed distributions for the SIRT6 KO cells with time. To define the distributions for short times we did not subtract the background in all the four quadrants (as was done for all the other but this calculation). In our case, when apoptosis was induced by starvation, the distributions were reasonably static. However, the directional movement might be observed in some other cases.

We conjecture that net directional movement of cell staining distributions might be observed in cases when apoptosis is induced by DNA damaging agents, and could represent a process of DNA degradation. Such information might be of interest to researchers studying dynamics of cell death in response to chemotherapeutic treatments.

However, before considering information derived from the distributions of the dye-staining intensities as definitive, the use of the logarithmic data presentation, which has been widely utilized in the literature [20,21], but is in fact entirely *ad hoc*, should be questioned. Functions that relate the fluorescence intensities to measures of various types of cell-membrane damage that enable dye attachment or penetration, should be used instead (such functions are likely time-dependent and may be different for the two axes), but they are not known presently. Specifically, our data (Fig. 4) suggest that cells undergoing apoptosis not only have much larger dye amounts than normal cells, but also have the processes of the intake on the two dyes more correlated. However, the definitiveness of the latter conclusion depends on the variables used as the *x* and *y* axes, because their rescaling might change the angles.

In summary, our computational model combined with experimental data obtained from time-lapse experiments can provide a novel approach to understanding apoptotic dynamics, as well as quantifiably measure the impact that genetic or environmental modulations have on the process of programmed cell death.

## Materials and Methods

All resources generated from this study will be or already are openly shared with the research community.

### *Cell culture*

Mouse embryonic fibroblasts were cultured in Dulbecco's Modified Eagle Medium (DMEM) supplemented with 10% fetal bovine serum and penicillin/streptomycin antibiotics. Several independent batches of primary cells were isolated from different mice and all demonstrated the described phenotypes. The identity of cells was verified visually and via SIRT6 immunoblotting with anti-SIRT6 antibodies.

### *Sodium-dodecyl-sulfate polyacrylamide gel electrophoresis and Immunoblotting*

Cells were lysed in Radioimmunoprecipitation assay (RIPA) buffer (50 mM Tris at pH 7.4, 150 mM NaCl, 1M EDTA, 0.25% deoxy chloric acid, and 1% NP-40) supplemented with protease inhibitor cocktail (Roche, Cat# 4693116001). The mixture was centrifuged and supernatant was taken. Protein levels were standardized using a Bradford protein assay. Protein was mixed with sodium dodecyl sulfate and electrophoresed in a 12% acrylamide gels. Proteins were transferred to a Polyvinylidene difluoride (PVDF) membrane (0.45uM) and the membrane was then immunoblotted with antibodies, diluted at concentrations recommend by the manufacturer, against the specific proteins being examined. In this study we used the following **Antibodies**: anti-SIRT6 (rabbit polyclonal, Sigma-Aldrich, Cat# S4322), anti-$\beta$-actin (mouse monoclonal, Abcam, Cat#ab8226).

### *Cellular Stresses*

For inducing cellular stress and apoptosis the following treatments were used: Rotenone (10 $\mu$M), proteasome inhibitor MG132 (10 $\mu$M), serum starvation (fetal bovine serum was withheld from culture media), or etoposide (20 $\mu$M). Survival of cells was measured using flow cytometry 12 hours later, after cells had been stained with the apoptotic markers – Annexin and propidium iodide.

### ***Flow cytometry***

Mouse embryonic fibroblasts were collected from wells using Trypsin digestion. Cells were washed in Phosphate Buffered Saline (PBS) and then suspended in 100 μL of 1X binding buffer (10 mM HEPES, 140 mM NaCl, 2.5 mM $CaCl_2$, pH 7.4) with 5 μL of Annexin conjugate and PI. After a 15-minute incubation another 400 μL of binding buffer was added and then at least 2,000 cells were analyzed using a 3 laser/ 8 color Beckton-Dickinson LSR II.

## Acknowledgements

S.L. and J.N. were in part supported by a grant from American Federation for Aging Research (AFAR, grant # 2015-030). S.L. received seed grant funding from the Cornell University Center for Vertebrate Genomics. J.N. was supported by a Glenn/AFAR Scholarship for Research in the Biology of Aging.

## Author contributions

S.D. performed computational modelling and wrote the initial draft of the manuscript; J.N. performed cell culture experiments; J.E.S. assisted in optimizing the computational model; V.P. nucleated the method of modelling, supervised modelling efforts, designed the study; S.L. designed the study, supervised biological cell culture experiments, supervised computational modelling, edited the manuscript.

## Conflict of interest

All authors declare no conflicts of interest.

## References

1. Johnstone RW, Ruefli AA, Lowe SW (2002) Apoptosis: a link between cancer genetics and chemotherapy. Cell 108 (2):153-164. doi:S0092867402006256 [pii]
2. Perry G, Nunomura A, Lucassen P, Lassmann H, Smith MA (1998) Apoptosis and Alzheimer's disease. Science 282 (5392):1268-1269
3. Liu Y, Song Y, Zhu X (2017) MicroRNA-181a Regulates Apoptosis and Autophagy Process in Parkinson's Disease by Inhibiting p38 Mitogen-Activated Protein Kinase (MAPK)/c-Jun N-Terminal Kinases (JNK) Signaling Pathways. Med Sci Monit 23:1597-1606. doi:900218
4. Delfino DV, Pozzesi N, Pierangeli S, Ayroldi E, Fierabracci A (2011) Manipulating thymic apoptosis for future therapy of autoimmune diseases. Curr Pharm Des 17 (29):3108-3119. doi:BSP/CPD/E-Pub/000582 [pii]
5. Muller PA, Vousden KH (2013) p53 mutations in cancer. Nat Cell Biol 15 (1):2-8. doi:10.1038/ncb2641
6. Sebastian C, Zwaans BM, Silberman DM, Gymrek M, Goren A, Zhong L, Ram O, Truelove J, Guimaraes AR, Toiber D, Cosentino C, Greenson JK, MacDonald AI, McGlynn L, Maxwell F, Edwards J, Giacosa S, Guccione E, Weissleder R, Bernstein BE, Regev A, Shiels PG, Lombard DB, Mostoslavsky R (2012) The histone deacetylase SIRT6 is a tumor suppressor that controls cancer metabolism. Cell 151 (6):1185-1199. doi:10.1016/j.cell.2012.10.047
7. Libert S, Guarente L (2013) Metabolic and neuropsychiatric effects of calorie restriction and sirtuins. Annu Rev Physiol 75:669-684. doi:10.1146/annurev-physiol-030212-183800
8. Lin SJ, Kaeberlein M, Andalis AA, Sturtz LA, Defossez PA, Culotta VC, Fink GR, Guarente L (2002) Calorie restriction extends Saccharomyces cerevisiae lifespan by increasing respiration. Nature 418 (6895):344-348. doi:10.1038/nature00829
9. Libert S, Zwiener J, Chu X, Vanvoorhies W, Roman G, Pletcher SD (2007) Regulation of Drosophila life span by olfaction and food-derived odors. Science 315 (5815):1133-1137. doi:1136610 [pii]
10. Mattison JA, Colman RJ, Beasley TM, Allison DB, Kemnitz JW, Roth GS, Ingram DK, Weindruch R, de Cabo R, Anderson RM (2017) Caloric restriction improves health and survival of rhesus monkeys. Nat Commun 8:14063. doi:10.1038/ncomms14063
11. Kanfi Y, Naiman S, Amir G, Peshti V, Zinman G, Nahum L, Bar-Joseph Z, Cohen HY (2012) The sirtuin SIRT6 regulates lifespan in male mice. Nature 483 (7388):218-221. doi:10.1038/nature10815
12. TenNapel MJ, Lynch CF, Burns TL, Wallace R, Smith BJ, Button A, Domann FE (2014) SIRT6 minor allele genotype is associated with >5-year decrease in lifespan in an aged cohort. PLoS One 9 (12):e115616. doi:10.1371/journal.pone.0115616
13. Van Meter M, Mao Z, Gorbunova V, Seluanov A (2011) SIRT6 overexpression induces massive apoptosis in cancer cells but not in normal cells. Cell Cycle 10 (18):3153-3158. doi:17435 [pii]
14. Pfister JA, Ma C, Morrison BE, D'Mello SR (2008) Opposing effects of sirtuins on neuronal survival: SIRT1-mediated neuroprotection is independent of its deacetylase activity. PLoS One 3 (12):e4090. doi:10.1371/journal.pone.0004090
15. Ouyang L, Shi Z, Zhao S, Wang FT, Zhou TT, Liu B, Bao JK (2012) Programmed cell death pathways in cancer: a review of apoptosis, autophagy and programmed necrosis. Cell Prolif 45 (6):487-498. doi:10.1111/j.1365-2184.2012.00845.x
16. Crowley LC, Marfell BJ, Scott AP, Waterhouse NJ (2016) Quantitation of Apoptosis and Necrosis by Annexin V Binding, Propidium Iodide Uptake, and Flow Cytometry. Cold Spring Harb Protoc 2016 (11):pdb prot087288. doi:10.1101/pdb.prot087288

17. Domanskyi S, Schilling JE, Gorshkov V, Libert S, Privman V (2016) Rate-equation modelling and ensemble approach to extraction of parameters for viral infection-induced cell apoptosis and necrosis. J Chem Phys 145 (9):094103. doi:10.1063/1.4961676
18. Khammanit R, Chantakru S, Kitiyanant Y, Saikhun J (2008) Effect of serum starvation and chemical inhibitors on cell cycle synchronization of canine dermal fibroblasts. Theriogenology 70 (1):27-34. doi:10.1016/j.theriogenology.2008.02.015
19. Yoshimoto N, Tasaki M, Shimanouchi T, Umakoshi H, Kuboi R (2005) Oxidation of cholesterol catalyzed by amyloid beta-peptide (A beta)-Cu complex on lipid membrane. J Biosci Bioeng 100 (4):455-459. doi:10.1263/jbb.100.455
20. Arsenio J, Kakaradov B, Metz PJ, Kim SH, Yeo GW, Chang JT (2014) Early specification of CD8+ T lymphocyte fates during adaptive immunity revealed by single-cell gene-expression analyses. Nat Immunol 15 (4):365-372. doi:10.1038/ni.2842
21. Hollville E, Martin SJ (2016) Measuring Apoptosis by Microscopy and Flow Cytometry. Curr Protoc Immunol 112:14 38 11-14 38 24. doi:10.1002/0471142735.im1438s112